\documentclass[epj,final]{svjour}
\usepackage{graphicx}
\usepackage[english]{babel}
\usepackage{amssymb,amsmath}
\newcommand{\bp}{{\bf p}}

\newcommand{\bk}{{\bf k}}
\newcommand{\beq}{\begin{equation}}
\newcommand{\eeq}{\end{equation}}

\begin{document}

\title{Repulsive polarons and itinerant ferromagnetism in strongly polarized Fermi gases}
\author{P.\ Massignan\inst{1,}\inst{2} \and G.\ M.\ Bruun\inst{3}}
\institute{ICFO-Institut de Ciencies Fotoniques, 08860 Castelldefels (Barcelona), Spain
\and F\'isica Teorica: Informaci\'o i Processos Qu\`antics, Universitat Aut\`onoma de Barcelona, 08193 Bellaterra, Spain
\and Department of Physics and Astronomy, University of Aarhus, DK-8000 Aarhus C, Denmark}
\date{Received: \today / Revised version: date}

\abstract{We analyze the properties of a single impurity immersed in a  Fermi sea. At positive energy and scattering lengths, we show that the system possesses a well-defined but metastable excitation, the repulsive polaron, and we calculate its energy, quasiparticle residue and effective mass.
From a thermodynamic argument we obtain the number of particles in the dressing cloud, illustrating the repulsive character of the polaron. 
Identifying the important 2- and 3-body decay channels, we furthermore calculate the lifetime of the repulsive polaron. The stability conditions for the formation of fully spin polarized (ferromagnetic) domains are then examined for a binary mixture of atoms with a general mass ratio. Our results indicate that mass imbalance lowers the critical interaction strength for phase-separation, but that  very short quasiparticle decay times will complicate the
 experimental observation of itinerant ferromagnetism.
 Finally, we present the spectral function of the impurity for various coupling strengths and momenta.
}

\maketitle

%%%%  %%%%  %%%%  %%%%

\section{Introduction}
Ultracold gases provide a powerful system to study many-body effects in quantum systems~\cite{RevMods}.
 In particular, mixtures of fermionic gases close to a Feshbach resonance represent  systems where the interactions may be tuned at will, 
 becoming strongly correlated in the unitarity limit~\cite{PolaronPapers}.
 Experimentally, it is possible to control % with great freedom 
 the number of atoms in various hyperfine states, the temperature and the dimensionality of the system.
 Moreover, heteronuclear Fermi-Fermi mixtures has recently become available in various laboratories~\cite{Taglieber08,Wille08,Tiecke10,Naik10,Fukuhara07%,Okano10
}, opening new exciting research possibilities.
 
 A two-component Fermi gas with equally populated components and attractive interactions gives rise to rich physics related to the cross-over between a BCS superfluid and a Bose-Einstein condensate (BEC). 
 In the opposite limit of extreme polarization  which corresponds to a few impurity atoms diluted in an otherwise ideal Fermi sea, one obtains instead a realization of Fermi polaron physics~\cite{Schirotzek09}.
 In three dimensions one can get a quantitative description of highly polarized systems using a one particle-hole wavefunction~\cite{PolaronPapers}. In lower dimensions, higher orders in this hole expansion are needed to take into account the increasing role of quantum fluctuations~\cite{Zoellner,Parish11,Giraud09,Leskinen10}.
 
 It is of great interest now to investigate the region of parameters where atoms experience repulsive interactions. Indeed, as one increases
  the repulsion between two distinguishable fermionic gases it is predicted that the system should undergo a transition from a mixed phase to a state characterized by spatially-separated, polarized domains.
  More than 70 years after the original proposal of Stoner \cite{Stoner33}, the transition towards itinerant ferromagnetism (IFM) is not  fully understood, nor conclusively demonstrated experimentally.
 Even though the first experimental results hinting at such transition have been recently reported~\cite{Jo}, their interpretation is still object of intense debate~\cite{Zhai09,Cui10,Pilati10,Pekker10}.  
  
In this paper we study 
%present a complete characterization of the fundamental excitations of a strongly-imbalanced Fermi gas with repulsive interactions, and we discuss how the intrinsic metastability of the relevant quasiparticles may affect the experimental observation of the instability towards the ferromagnetic state.
 %To achieve this goal, we study
  a single $\downarrow$ impurity atom immersed in a Fermi sea of $\uparrow$ atoms in 3D using a hole-expansion. The $\uparrow-\downarrow$ interactions are assumed to be tunable by means of a broad Feshbach resonance.
The spectral function of the impurity is shown to exhibit two distinct quasiparticle peaks. The high energy peak corresponds to 
the lowest repulsive (scattering) state of a Feshbach resonance on the BEC side with positive scattering length $a$.
 We refer to this  quasiparticle state as the repulsive polaron, and we calculate its energy, effective mass, and residue as a function of interaction strength and impurity mass.
 A good level of confidence over our findings is ensured by analytic checks in the limits of weak coupling and  infinite impurity mass. 
  We also calculate the number of majority atoms in the screening cloud of both the repulsive and the attractive polarons, 
 by means of a thermodynamic argument. The decay rate of repulsive polarons is then analyzed by considering the relevant 2- and 3-body decay processes.   
    We switch to considering the stability  conditions for  a perfectly ferromagnetic state, showing that mass imbalance favors
     ferromagnetism.
    However, we find that the corresponding quasiparticle lifetimes are too short to permit the experimental observation of ferromagnetism in universal Fermi-Fermi mixtures.  
    Finally, we plot the spectral function of the impurity as a function of momentum and interaction strength, and we discuss its connection to radio frequency (rf) spectroscopy.

%%%%  %%%%  %%%%  %%%%

\section{Formalism}
An accurate characterization of attractive polarons can be obtained in terms of the following variational ansatz~\cite{Chevy}
\begin{equation}
|\psi\rangle= \phi_0 a_{{\mathbf p}\downarrow}^\dagger|{\rm FS}\rangle + 
\sum_{q<k_F<k}\phi_{{\bf k},{\bf q}} a_{{\mathbf p}+{\mathbf q}-{\mathbf k}\downarrow}^\dagger
a^{\dag}_{{\bf k}\uparrow}\,a_{{\bf q}\uparrow}|{\rm FS}\rangle,
\label{ansatz}
\end{equation}
where $a_{{\bf k}\sigma}$ annihilates a fermion of species $\sigma$ with momentum ${\mathbf k}$. 
The ansatz describes an impurity $\downarrow$  with momentum ${\mathbf p}$
in a Fermi sea $|{\rm FS}\rangle$ of $\uparrow$ particles,  dressed  by  particle-hole pairs.
The energy and effective mass given by this variational ansatz compare very well
 with both Monte-Carlo (MC) simulations \cite{Prokofev08} and experiments \cite{Schirotzek09}.

Diagrammatically, the polaron properties are given in terms of the retarded self-energy $\Sigma$ of a single impurity of mass $m_{\downarrow}$ in
a Fermi sea of particles with mass $m_{\uparrow}$ and density $n_\uparrow=k_F^3/6\pi^2$.
The minimization of the energy based on the ansatz (\ref{ansatz}) is identical to a diagrammatic calculation of the polaron properties within the ladder
 (or "forward-scattering") approximation \cite{Combescot07}.
 At $T=0$, the latter yields
\begin{equation}
\Sigma(\mathbf{p},E)=\int d^3\check{q}
\frac{\theta(k_F-q)}{\frac 1 {T_0
}-\int d^3\check k\left(\frac{2m_\mathrm{r}}{k^{2}}+
\frac{\theta(k-k_F)}{E+i0^{+}-\Delta\epsilon_{\mathbf{p+q},\mathbf{k}}}\right)}
\label{eq:SelfEnergy}
\end{equation}
with  $m_\mathrm{r}=m_{\uparrow}m_{\downarrow}/(m_{\uparrow}+m_{\downarrow})$ the reduced mass of a $\uparrow\downarrow$ pair
and $T_0
=2\pi a/m_\mathrm{r}$ the zero energy vacuum scattering matrix. 
We have defined $\Delta\epsilon_{\mathbf{p+q},\mathbf{k}}=\epsilon_{\uparrow k}-\epsilon_{\uparrow q}+\epsilon_{\downarrow\mathbf{p+q-k}}$ with  $\epsilon_{\sigma k}=k^2/2m_\sigma$, 
 $d^3\check k=d^3k/(2\pi)^3$ and set $\hbar=1$. 

\begin{figure}
\begin{center}
\includegraphics[width=\columnwidth]{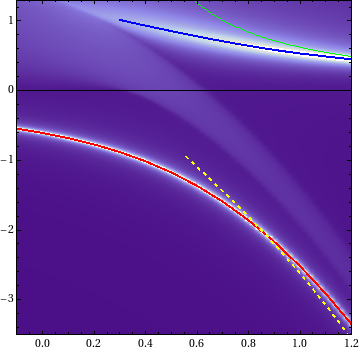}
\end{center}
\caption{Contourplot of the (square root of the) spectral function $A_\downarrow(\mathbf{p}=0,E)$ of an impurity in a Fermi sea with $m_\downarrow=m_\uparrow$. The x-axis is $1/k_Fa$, and the y-axis is $E/\epsilon_F$ with $\epsilon_F=k_F^2/2m$. The solid lines depict the polaron energies given by Eq.\ (\ref{qpEnergies}), and the 
dashed line is the perturbative expression of the molecule energy given by Eq.\ (\ref{muMolec}). The light-blue area between the two polaronic branches is the molecule-hole continuum.
}
\label{figSpectralFunctionSquareRoot}
\end{figure}

In Fig.\ \ref{figSpectralFunctionSquareRoot} we plot the spectral function 
\beq
A_\downarrow(\bp,E)=-2\mathrm{Im}\frac{1}{E+i0^+-\epsilon_{\downarrow p}-\Sigma(\mathbf{p},E)}
\label{Spectral}
\eeq
as a function of interaction strength $1/k_Fa$ and energy $E$ for an impurity with momentum $\bp=0$ for the equal masses case $m_\downarrow=m_\uparrow$. 
%%%%figSpectralFunction%%%%%%%%%%%%%%%%%%%%%%%%%%%%%%%%%%%%%%%%%%%%%%%%%%%%%%%%%%%%%%%%%%%%
%%%%%%%%%%%%%%%%%%%%%%%%%%%%%%%%%%%%%%%%%%%%%%%%%%%%%%%%%%%%%%%%%%%%%%%%
At small momenta, the spectral function is strongly peaked at energies satisfying
\begin{equation}
E={\rm Re}[\Sigma(\mathbf{p}=0,E)].
\label{qpEnergies}
\end{equation}
For positive scattering lengths, this self-consistent equation has two solutions.
The lower branch at energy $E_-<0$ corresponds to the well known attractive polaron
 investigated experimentally in Ref.\ \cite{Schirotzek09}. Although the attractive polaron is undamped
 in the ladder approximation,  it is in fact unstable towards decay into 
a molecule, two holes and a particle for $1/k_Fa\gtrsim 0.9$~\cite{Prokofev08,Punk}. 
The energy of a $\bp=0$ molecule in the deep BEC limit $1/k_Fa\gg 1$ is
\beq
E_M=-\frac{1}{2m_r a^2}-\epsilon_F+\frac{2\pi a_3}{m_3}n_\uparrow,
\label{muMolec}
\eeq
with $m_3$ and $a_3$ the atom-dimer $(\uparrow-\uparrow\downarrow)$ reduced mass and scattering length \cite{Petrov03}.
To describe the decay into the molecule-hole continuum, it is necessary to go beyond the ladder approximation and include more particle-hole excitations~\cite{Bruun10}. Including 
such processes moves the threshold of the molecule-hole continuum, visible in Fig.\ \ref{figSpectralFunctionSquareRoot} as a broad peak between the two polaronic branches, down in energy all the way to the yellow dashed line, yielding a non-zero width to the attractive polaron peak for $1/k_Fa\gtrsim 0.9$.

The upper peak of the spectral function at energy $E_+>0$ corresponds to the repulsive polaron. 
This excited state has a finite width coming from the decay into the lower-lying excitations. The decay rate $\Gamma$ will be analyzed in detail in Sec.\ \ref{sec:Decay}.
The repulsive polaron is well defined as long as its decay rate is much smaller than its energy, $\Gamma/E_+\ll1$.
In the rest of the paper, we mainly focus on  the quasiparticle properties of the repulsive polaron. 
 Its quasiparticle residue is defined as
\beq
Z_+=\left\{1-\mathrm{Re}\left[\frac{\partial\Sigma(\bp=0,E_+)}{\partial E}\right]_{E_+}\right\}^{-1}
\eeq
and its effective mass as
\beq
m_{+}^*=\frac{m_\downarrow}{Z}
\left\{1+\mathrm{Re}\left[\frac{\partial\Sigma(\bp,E_+)}{\partial\epsilon_{\downarrow p}}\right]_{p=0}\right\}^{-1}.
\eeq

%%%%  %%%%  %%%%  %%%%

\section{Energy, residue and effective mass of the repulsive polarons}
The energy of an infinitely massive impurity interacting repulsively with the surrounding Fermi gas may be calculated by means of the Fumi theorem \cite{Fumi,Combescot07,Zoellner}. We find
\beq
\frac{E_+}{\epsilon_F}=-\frac {1}{\pi}\left\{\left(1+y^2\right)\left[-\frac{\pi}{2}+\arctan(y)\right]+y\right\},
\label{Fumi}
\eeq
where $y=1/k_Fa$ which is depicted as a red dash-dotted line in Fig.\ \ref{figEnergy}. At resonance, an infinitely massive impurity has the energy $\epsilon_F/2$.

For a generic mass ratio, the polaron properties may be calculated analytically in the weak coupling limit \mbox{$k_F|a|\ll 1$} as an expansion in powers of $k_Fa$~\cite{Bishop73}. 
One finds for the polaron energy
\begin{equation}
\frac{E_+}{\epsilon_F}=\frac{m_\uparrow}{m_\mathrm{r}}\left[\frac{2}{3\pi}k_Fa+F(\alpha)(k_Fa)^2+O[(k_Fa)^3]\right],
\label{muDownAnalytic}
\end{equation}
with $\alpha=(m_\downarrow-m_\uparrow)/(m_\downarrow+m_\uparrow)$ and
\begin{equation}
F(\alpha)=\frac{1-\alpha}{4\pi^2\alpha^2}\left[(1+\alpha)^2\log\left(\frac{1+\alpha}{1-\alpha}\right)-2\alpha\right].
\end{equation}
We stress here that the expansion is valid for both $k_Fa\rightarrow 0^\pm$, i.e., for the repulsive as well as the attractive polaronic branches.
For equal masses $\lim_{\alpha\rightarrow 0}F(\alpha)=1/\pi^2$, and one recovers the known result
\begin{equation}
\frac{E_+}{\epsilon_F}=\frac{4}{3\pi}k_Fa+\frac{2}{\pi^2}(k_Fa)^2+O[(k_Fa)^3].
\label{muDownBishopBalanced}
\end{equation}

In the strongly-interacting regime, we calculate the energy $E_+$ of the repulsive polaron  numerically from Eq.\ ({\ref{qpEnergies}). Our results are shown in Fig.\ \ref{figEnergy} for various mass ratios, corresponding to 
experimentally relevant mixtures of   $^6$Li, and $^{40}$K atoms~\cite{Taglieber08,Wille08,Tiecke10,Naik10}. For $k_Fa\ll 1$,
 we find good agreement with both the analytic result Eq.\ (\ref{muDownBishopBalanced}), and with the MC results of Ref.\ \cite{Pilati10} for an attractive square well. 
 For larger  values of $k_Fa$, the diagrammatic results lie below the MC results. The discrepancy may be explained  
 by the fact that the MC results, obtained through a variational approach, are strictly an upper bound to the energy. 
 Also, it could be due to particle-hole processes ignored by the ladder approximation. We  note that in the limit of a heavy impurity, our results approach the exact result (\ref{Fumi}) providing further evidence of the accuracy of the ladder approximation in the limit of strong polarization. 
 The energy  for equal masses agrees with the one found in Ref.\ \cite{Cui10}, which obtained the real part of the energy using  the variational ansatz (\ref{ansatz}).
 From Fig.\ \ref{figEnergy}, we see that  quasiparticles formed by light impurities have larger energies.

%%%%figEnergy%%%%%%%%%%%%%%%%%%%%%%%%%%%%%%%%%%%%%%%%%%%%%%%%%%%%%%%%%%%%%%%%%%%%
\begin{figure}
\begin{center}
\includegraphics[width=\columnwidth]{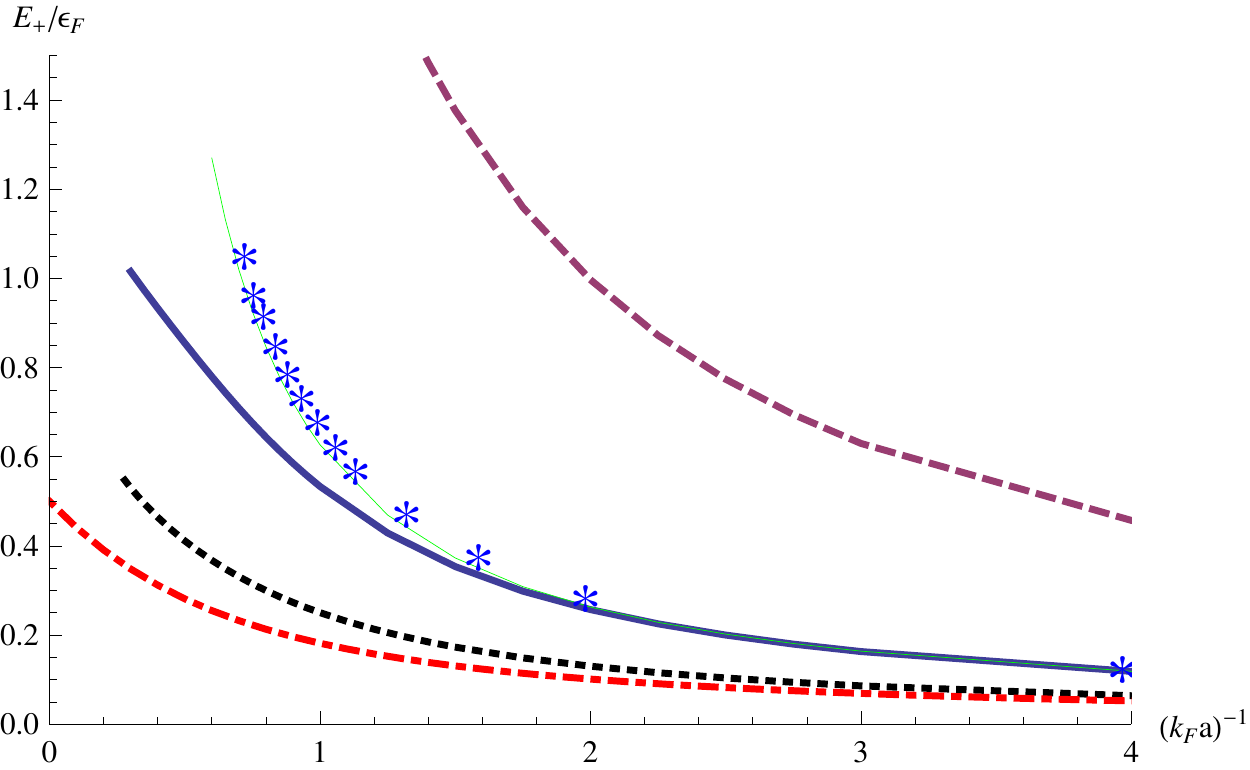}
\end{center}
\caption{Energy of the repulsive polaron. From top to bottom (thick lines), the impurity becomes increasingly heavier: $m_\downarrow/m_\uparrow=6/40,\,1,\,40/6,\,\infty$. The infinite mass result is obtained from (\ref{Fumi}). For equal masses we plot also the weak-coupling limit Eq.\ (\ref{muDownBishopBalanced}) (thin green line), and the MC results for square well potentials of Ref.\ \cite{Pilati10} (symbols).}
\label{figEnergy}
\end{figure}
%%%%%%%%%%%%%%%%%%%%%%%%%%%%%%%%%%%%%%%%%%%%%%%%%%%%%%%%%%%%%%%%%%%%%%%%

%%%%figMstarAndZ%%%%%%%%%%%%%%%%%%%%%%%%%%%%%%%%%%%%%%%%%%%%%%%%%%%%%%%%%%%%%%%%%%%%
\begin{figure}
\begin{center}
\includegraphics[width=\columnwidth]{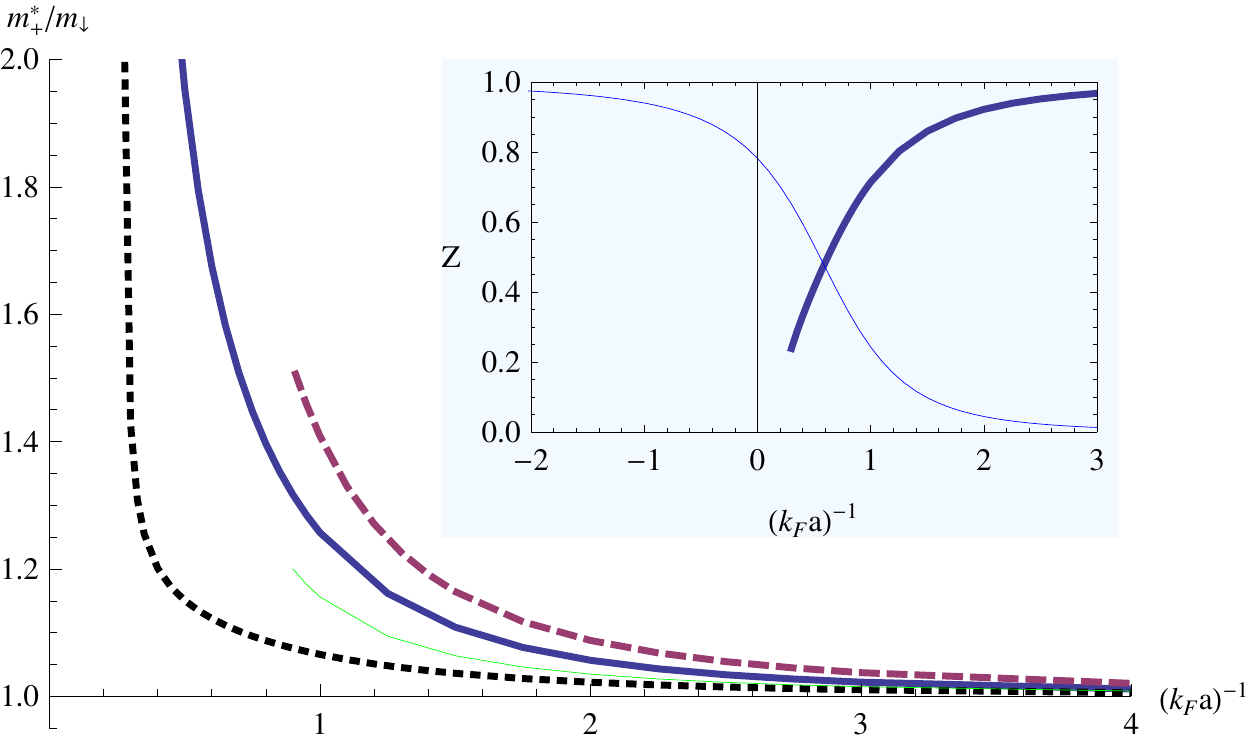}
\end{center}
\caption{Effective mass of the repulsive polaron. From top to bottom, $m_\downarrow/m_\uparrow=6/40,\,1,\,40/6$. The thin green line is the weak-coupling result for the equal masses case,  Eq.\ (\ref{mEffBishop}). Inset: quasiparticle residue Z for the attractive (thin) and repulsive (thick) polaron for the case of equal masses.}
\label{figMeffWithZInInset}
\end{figure}
%%%%%%%%%%%%%%%%%%%%%%%%%%%%%%%%%%%%%%%%%%%%%%%%%%%%%%%%%%%%%%%%%%%%%%%%

The effective mass $m_{+}^*$, and the quasiparticle residue $Z_{+}$ of the repulsive polaron are plotted in Fig.\ \ref{figMeffWithZInInset}. 
 In the BEC limit we have $Z_{+}\rightarrow 1$, whereas it rapidly decreases in the strongly-interacting regime $1/k_F|a|\ll 1$.
 Note that the residue $Z_{-}$ of the attractive polaron, which is also plotted in the inset of Fig.\ \ref{figMeffWithZInInset}, has the opposite behavior: it approaches unity in the weak coupling BCS limit $k_Fa\rightarrow 0_-$
 , and it decreases rapidly in the BEC limit $k_Fa\rightarrow 0_+$. The sum of the two residues is very close to unity for all interaction strengths.
 The effective mass  of the positive polaron reduces to the bare impurity mass $m_\downarrow$ in the BEC limit, and it rises rapidly for $1/k_Fa\rightarrow 0$ as the polaron gets increasingly dressed by particle-hole pairs, eventually becoming infinite in the vicinity of the resonance.
In the weakly-interacting limit, one has an analytic expression for the effective mass for a generic mass ratio~\cite{Bishop73},
\beq
\frac{m_\downarrow}{m_{+}^*}=1-\frac{2}{3\pi^2}\left[\frac{1}{\alpha}-\frac{(1-\alpha)^2}{2\alpha^2}\log\left(\frac{1+\alpha}{1-\alpha}\right)\right](k_Fa)^2+\ldots .
\label{mEffBishop}
\eeq
We see in
Fig.\ \ref{figMeffWithZInInset} that the numerical results approach this analytical expression in the limit $(k_Fa)^{-1}\gg 1$.

%%%%  %%%%  %%%%  %%%%

 \section{Number of particles in the dressing cloud}
To obtain a better understanding of the polaron wave function, we now calculate the number $\Delta N$ of majority atoms in its dressing cloud using a thermodynamic argument \cite{Massignan05,Zoellner}:
 $\Delta N$ is obtained by requiring that the density of the majority atoms  far away from the impurity  remains unchanged when adding one impurity.
 This corresponds to a constant chemical potential $\mu_\uparrow$ for the majority atoms giving
 \begin{equation}
\delta\mu_\uparrow=
\frac{\partial^{2}\varepsilon}{\partial n_\uparrow\partial n_\downarrow}+\frac{\partial^{2}\varepsilon}{(\partial n_\uparrow)^{2}}\Delta N=0
\end{equation}
where $\varepsilon(n_\uparrow,n_\downarrow)$ is the energy density of a gas, and $n_\uparrow$ and $n_\downarrow$ are the particle densities.
Solving for $\Delta N$, we find
\begin{equation}
\Delta N=-\left(\frac{\partial\mu_\downarrow}{\partial n_\uparrow}\right)_{n_\downarrow}/\left(\frac{\partial\mu_\uparrow}{\partial n_\uparrow}\right)_{n_\downarrow}
%=-\left(\frac{\partial\mu_\downarrow}{\partial \mu_\uparrow}\right)_{n_\downarrow}
\approx-\left(\frac{\partial\mu_\downarrow}{\partial\epsilon_{F}}\right)_{n_\downarrow},
\end{equation}
where we have used $\mu_\uparrow\approx\epsilon_{F}$  in the last equality.

Numerical results for $\Delta N$ obtained from $-\partial E/\partial\epsilon_{F}$ for the case of equal masses are presented in Fig.\ \ref{figDeltaNBalanced} for both the repulsive and the attractive polaron. We see that $\Delta N<0$ for the repulsive polaron, which corresponds 
to the majority atoms being pushed away as expected. The effect increases with increasing interaction strength. 
Likewise, the attractive polaron pulls in the majority atoms giving $\Delta N>0$.
The ansatz (\ref{ansatz}) becomes increasingy accurate 
for both polaronic states as $k_Fa\rightarrow 0$ and $E_{\pm}\rightarrow 0$.
In this limit one can even obtain an analytic expression, differentiating Eq.\ (\ref{muDownBishopBalanced}) to obtain
\beq
\Delta N=-\frac{2}{\pi}k_Fa-\frac{4}{\pi^2}(k_Fa)^2+\ldots
\label{DeltaNanalytic}
\eeq

 We see however that 
the number of majority atoms dressing the attractive polaron does not increase mono-tonously with increasing attraction going towards the deep BEC limit where $E_{-}\rightarrow -\infty$. 
This surprising result is due to the breakdown of the ansatz (\ref{ansatz}) for the attractive polaron in the BEC limit.
 Here the real ground state is the molecule with energy given by  Eq.\ (\ref{muMolec}). 
Taking its derivative, we find
\begin{equation}
\Delta N_M=-\frac{\partial \mu_M}{\partial\epsilon_{F}}=1-(k_Fa)\frac{a_3}{a}\frac{r+2}{\pi(r+1)},
\label{DeltaNmolec}
\end{equation}
with $r=m_\downarrow/m_\uparrow$ which is also plotted in Fig.\ \ref{figDeltaNBalanced}. So even though the  
wave function (\ref{ansatz}) yields an energy rather close to the correct molecule 
energy in the BEC limit, it predicts a qualitatively wrong result for the associated number of particles in the dressing cloud, of approximatively 0.5. 
On the contrary, $\Delta N$ for the molecule correctly approaches unity.
A similar but even stronger effect occurs in 2D~\cite{Zoellner}.

%%%%%%%%%%%%%%%%%%%%%%%%%%%%%%%%%%%%%%%%%%%%%%%%%%%%%%%%%%%%%%%%%%%%%%%%

\begin{figure}
\begin{center}
\includegraphics[width=\columnwidth]{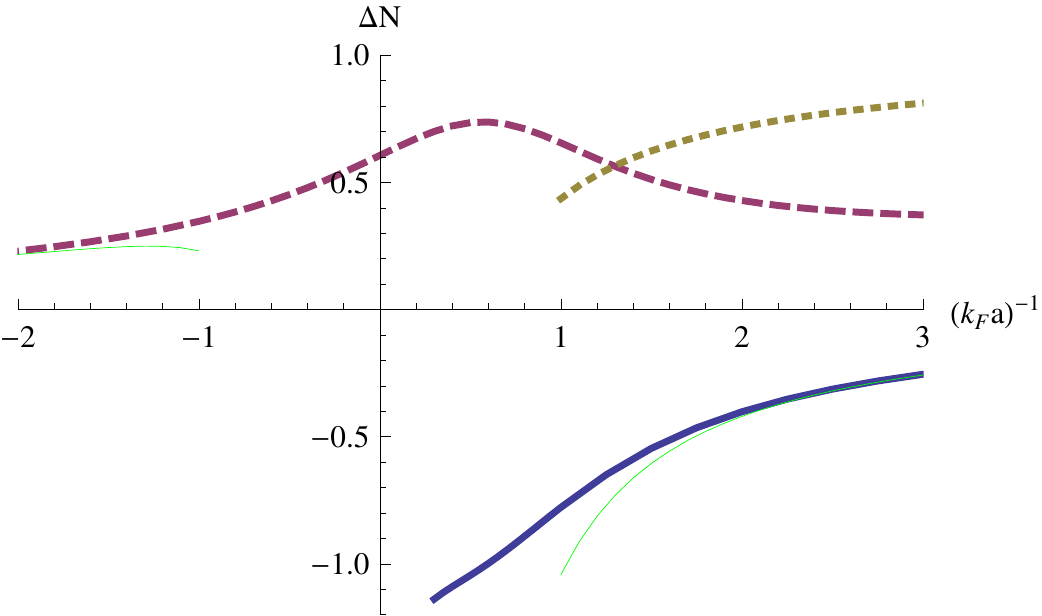}
\end{center}
\caption{Number $\Delta N$ of particles in the dressing clouds of the attractive (dashed line) and repulsive (solid line) polaron for 
 $m_{\downarrow}=m_{\uparrow}$. The thin green and dotted yellow lines are the analytic limits for the polarons and the molecule, respectively Eqs.\ (\ref{DeltaNanalytic}) and (\ref{DeltaNmolec}).}
\label{figDeltaNBalanced}
\end{figure}
%%%%%%%%%%%%%%%%%%%%%%%%%%%%%%%%%%%%%%%%%%%%%%%%%%%%%%%%%%%%%%%%%%%%%%%%

%%%%  %%%%  %%%%  %%%%

\section{Decay rate of the repulsive polaron}
\label{sec:Decay}
The calculation of the lifetime of the repulsive polaron is a complicated task, since it requires the self-energy to contain the correct low energy states 
into which the polaron can decay. In the quasiparticle picture,  these states are  the attractive polaron and the molecule-hole continuum. 
Neither of these decay channels are included in the non-selfconsistent ladder approximation  given by (\ref{eq:SelfEnergy}), which only contains the non-interacting $\downarrow$ Green's function. This could be remedied by using self-consistent Green's 
functions in the self-energy.
% which also has to include processes involving two majority holes not contained in the ladder approximation
%. Instead of 
%this a computationally intensive method, 
We take here a more pragmatic and much simpler approach, which contains the relevant physics for the decay; it is based on pole expansions of the interacting impurity and molecule Green's functions. 

\textbf{2-body decay:} In the simplest process, a repulsive polaron decays into an attractive polaron, scattering a majority particle  out of the Fermi sea.
To describe this 2-body decay channel, we use the fact that the impurity Green's function can be approximated as
\beq
 G^{\pm}_{\downarrow}(\bp,\omega)\sim\frac{Z_\pm}{\omega-E_\pm-p^2/2m_{\pm}^{*}}
\label{polaronsPoleExpansion}
\eeq
in the vicinity of each  pole $E_{\pm}$.
Replacing the bare $\downarrow$ Green's function with $ G^{-}_{\downarrow}$ inside the self-energy given by Eq.\ (\ref{eq:SelfEnergy}), we find a 
zero momentum repulsive  to attractive polaron decay rate  given by 
\begin{equation}
\Gamma_{PP}=-Z_+\mathrm{Im}[\Sigma^{-}(\mathbf{p}=0,E_+)],
\label{2bodyDecay}
\end{equation}
with
\begin{equation}
\Sigma^{-}(\mathbf{p},E)=\int d^3\check{q}
\frac{\theta(k_F-q)}{\frac 1 {T_0
}-Z_{-}\int d^3\check k\left(\frac{2m_\mathrm{r}}{k^{2}}+
\frac{\theta(k-k_F)}{E+i0^{+}-\Delta\epsilon^*_{\mathbf{p+q},\mathbf{k}}}\right)}.
\label{PPdecay}
\end{equation}
Here  $\Delta\epsilon^*_{\mathbf{p+q},\mathbf{k}}=E_{-}+\epsilon_{\uparrow k}-\epsilon_{\uparrow q}+\epsilon^*_{\downarrow\mathbf{p+q-k}}$ with  $\epsilon^*_{\downarrow k}=k^2/2m^*_{-}$ the kinetic energy of the attractive polaron.  In the BEC limit, we can expand (\ref{PPdecay})  in $k_Fa$ obtaining 
\begin{gather}
 \Gamma_{PP}=\pi T_0^2 Z_+ Z_-\int_{q<k_F<k} d^3\check{q}d^3\check{k}
 \delta(\Delta E+\epsilon_{\uparrow q}-\epsilon_{\uparrow k}-\epsilon^*_{\downarrow {\mathbf q}-{\mathbf k}})\nonumber\\
 =Z_+ Z_{-}\frac{2}{3\pi}\sqrt{\frac{m_\uparrow (m_r^*)^3}{m_r^4}}\sqrt{\frac{\Delta E_{PP}}{\epsilon_F}}(k_Fa)^2\epsilon_F.
\label{2body}
\end{gather}
We have defined $1/m_r^*=1/m_\uparrow+1/m_-^*$, $\Delta E_{PP}=E_+-E_-$,  and used that $\Delta E_{PP}\gg\epsilon_F$. Since in the BEC limit $\Delta E_{PP}\simeq 1/m_ra^2$, the 
decay rate into the lower polaron scales as $Z_- (k_Fa) (m_r^*)^{3/2}$. Note that Eqs.~(\ref{PPdecay})-(\ref{2body}) should be used only when the attractive polaron 
is a well-defined excitation with positive effective mass, i.e.\ for $(k_Fa)^{-1}\lesssim3$ for equal masses.
For simplicity, both analytical and numerical results are plotted in Fig.~\ref{figDecayAnalytic} assuming $m_r^*=m_r$.

\textbf{3-body decay:} The upper polaron can also decay via a more complex process, forming a molecule with one $\uparrow$ particle, and 
scattering another out of the Fermi sea. This  3-body channel can be described using self-energy terms which contain two holes and a molecule propagator 
 approximated by the pole expansion 
\beq
G_{M}(\bp,\omega)\sim\frac{Z_M}{\omega-E_M-p^2/2M^*}.
\label{moleculePoleExpansion}
\eeq
From a calculation analogous to the one presented in Ref.~\cite{Bruun10}, we find a zero-momentum polaron-to-molecule decay rate given by
\beq
 \frac{\Gamma_{PM}}{\epsilon_F}=\frac{32k_Fa}{45\pi^3}\frac{Z_{+}^3 Z_{M}}{(m^*_{+})^2\sqrt{m_\uparrow}}\left(1+\frac{m_\uparrow}{M}\right)^{3/2}\left(\frac{k_F}{\sqrt{2\Delta E_{PM}}}\right)^{5},
 \label{3bodyDecay}
\eeq
where $\Delta E_{PM}=E_+-E_M$. In the  BEC limit where the molecule energy is  $-1/2m_ra^2$, one finds that the three-body decay scales as $(k_Fa)^6$.
%, in accord with Ref.\ \cite{Petrov03} \textbf{Are you sure?}.

In Fig.~\ref{figDecayAnalytic} we plot the decay rates for a zero momentum repulsive polaron via the 2- and 3-body processes given by Eqs.~(\ref{2bodyDecay}) and (\ref{3bodyDecay}). As expected, the 2-body decay  is much faster than the 3-body decay~\cite{Sadeghzadeh11}. We see that the perturbative result (\ref{2body}) agrees with the full numerical calculation in the BEC limit.  
The decay rate obtained from the self-energy (\ref{eq:SelfEnergy}) containing bare $\downarrow$ propagators, $\Gamma_{\rm bare}=-Z_+ \mathrm{Im}[\Sigma(\mathbf{p}=0,E_+)]$, is also plotted. Although $\Gamma_{\rm bare}$ does not 
contain the correct physics to describe the decay of the repulsive polaron, we see that it yields a damping rate which 
is comparable to the correct one. 
%%%%figAnalyticDecayRates%%%%%%%%%%%%%%%%%%%%%%%%%%%%%%%%%%%%%%%%%%%%%%%%%%%%%%%%%%%%%%%
\begin{figure}
\begin{center}
\includegraphics[width=\columnwidth]{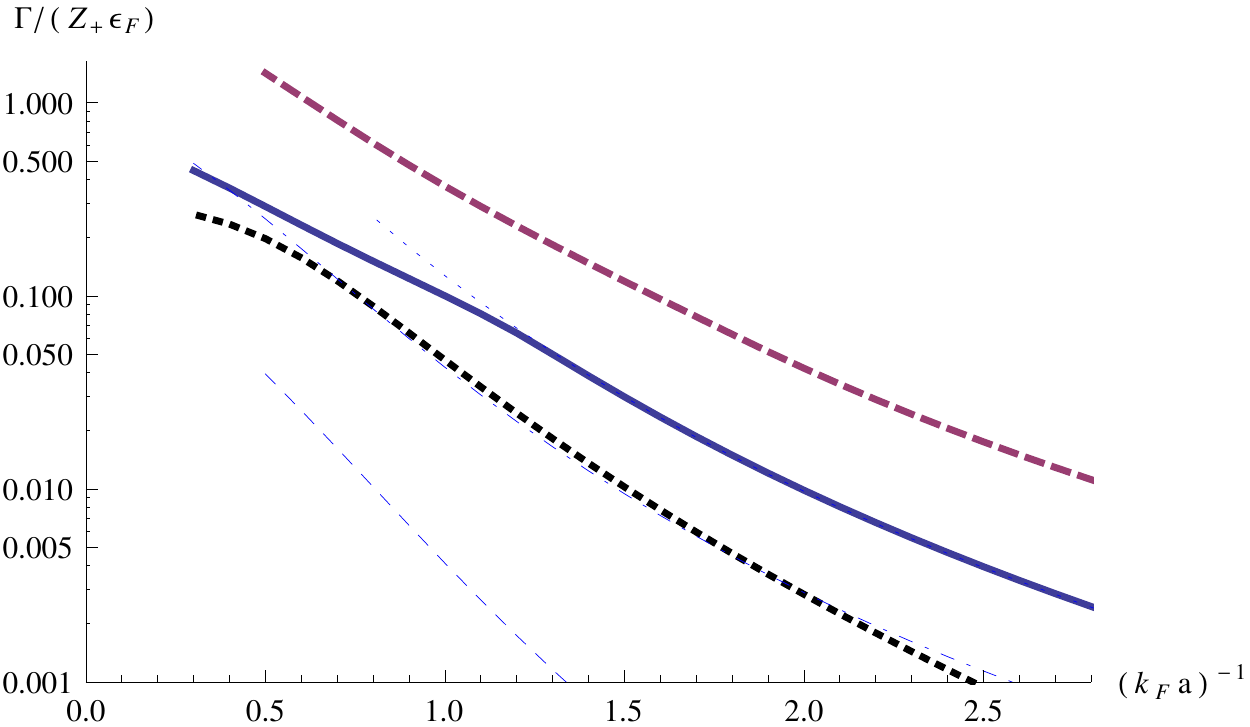}
\end{center}
\caption{Decay rate $\Gamma$ of the upper polaron. Thick lines represent the dominant two-body contribution $\Gamma_{PP}$, given by Eq.~(\ref{2bodyDecay}), for mass ratios $r=6/40$ (dashed), 1 (continuous), 40/6 (dotted). For equal masses we plot also as thin blue lines: the weaker three-body recombination process into atom+molecule, Eq. (\ref{3bodyDecay}) (dashed), the decay rate into bare impurities $\Gamma_{\rm bare}$ (dash-dotted), and the BEC limit of $\Gamma_{PP}$, Eq.~(\ref{2body}) (dotted).}
\label{figDecayAnalytic}
\end{figure}
%%%%%%%%%%%%%%%%%%%%%%%%%%%%%%%%%%%%%%%%%%%%%%%%%%%%%%%%%%%%%%%%%%%%%%%%

The decay rate is small compared to $\epsilon_F$ for  $1/k_Fa\gg 1$,
 showing that the repulsive polaron is a well-defined quasiparticle  in the BEC limit. 
 Approaching unitarity $\Gamma$ rapidly increases, and  eventually the repulsive polaron becomes ill-defined.
 % when $\Gamma\gtrsim E_{+}$.
  This behavior is reflected in the effective mass and the quasiparticle residue, which respectively diverge and vanish
   in this region. From Figs.\ \ref{figEnergy}-\ref{figDecayAnalytic}, we conclude that the effects of the interaction increase with decreasing impurity mass. 
This is as expected, since the interaction enters in Eq.~(\ref{eq:SelfEnergy}) in the combination $2\pi a/m_\mathrm{r}$, implying
 that lighter particles are scattered more effectively.

%%%%  %%%%  %%%%  %%%%

\section{Stability of the itinerant ferromagnetic phase}
In presence of strong repulsive interactions, the mean field argument of Stoner indicates that a two-component Fermi gas should undergo a transition to a state characterized by spin polarized domains~\cite{Stoner33}. This transition is  not fully understood  and has not yet been realized
 experimentally in an unambiguous way.
  The case of equal masses and populations has been analyzed in quite some detail theoretically.  Predictions for the critical interaction in this case range from the mean field result $k_Fa=\pi/2$, to the values of $k_Fa\simeq1.02$ and $0.8$ obtained respectively by second order perturbation theory \cite{Duine05} and by variational Monte Carlo ~\cite{Pilati10}. Recently,  experimental indications of a phase transition to a ferromagnetic state  were  reported in an atomic gas for $k_Fa\gtrsim1.9$~\cite{Jo} which 
  has caused much interest.  One complicating factor is that the itinerant ferromagnetic phase (IFM) competes with a pairing instability~\cite{Pekker10}.

 We discuss here the conditions for a fully ferromagnetic phase consisting of domains with complete spin polarization for a binary mixture of atoms with 
 general mass ratio.  Ferromagnetism of mass-imbalanced mixtures has recently been examined in the 
  mean-field approximation valid for weak coupling~\cite{Keyserlingk}. 
 Here, we use the results obtained in the previous sections, which should be accurate even for strong interactions,
 to analyze the stability of the fully separated phase.
  In this  phase,
   all $\uparrow$ particles reside in the phase $(\uparrow)$ with Fermi energy $\epsilon_{F\uparrow}$ and all $\downarrow$ particles in the 
   phase $(\downarrow)$ with Fermi energy $\epsilon_{F\downarrow}$. 
At equilibrium, the pressure of the two phases must be the same: $P_{(\uparrow)}=P_{(\downarrow)}$. 
Since the two phases are ideal gases, we have $P_{(\sigma)}=2n_\sigma\epsilon_{F\sigma}/5$ with $n_\sigma=(2m_\sigma \epsilon_{F\sigma})^{3/2}/6\pi^2$ the density. 
The equilibrium pressure condition therefore gives  $(k_{\uparrow F}/k_{\downarrow F})^5=m_\uparrow/m_\downarrow$. 
The  fully phase separated state  
is  stable if it is energetically unfavorable to move a particle across the phase boundary, i.e.\  if 
\beq
\epsilon_{F\uparrow}<E_{+}^{(\uparrow)} \quad \text{and}  \quad \epsilon_{F\downarrow}<E_{+}^{(\downarrow)},
\eeq
where  $E_{+}^{(\uparrow)}$ is the energy of a single $\uparrow$ particle in the repulsive polaron state in the $(\downarrow)$ phase, and likewise for $E_{+}^{(\downarrow)}$. Using the equilibrium pressure condition, these conditions can be written as 
\beq
\frac{E_{+}^{(\uparrow)}}{\epsilon_{F\downarrow}}\left(\frac{m_\uparrow}{m_\downarrow}\right)^{3/5}>1 \quad \text{and} \quad
\frac{E_{+}^{(\downarrow)}}{\epsilon_{F\uparrow}}\left(\frac{m_\downarrow}{m_\uparrow}\right)^{3/5}> 1.
\label{IFMconditions}
\eeq
%%%%figPowerThreeFifths%%%%%%%%%%%%%%%%%%%%%%%%%%%%%%%%%%%%%%%%%%%%%%%%%%%%%%%%%%%%%%%%%%%%
\begin{figure}
\begin{center}
\includegraphics[width=\columnwidth]{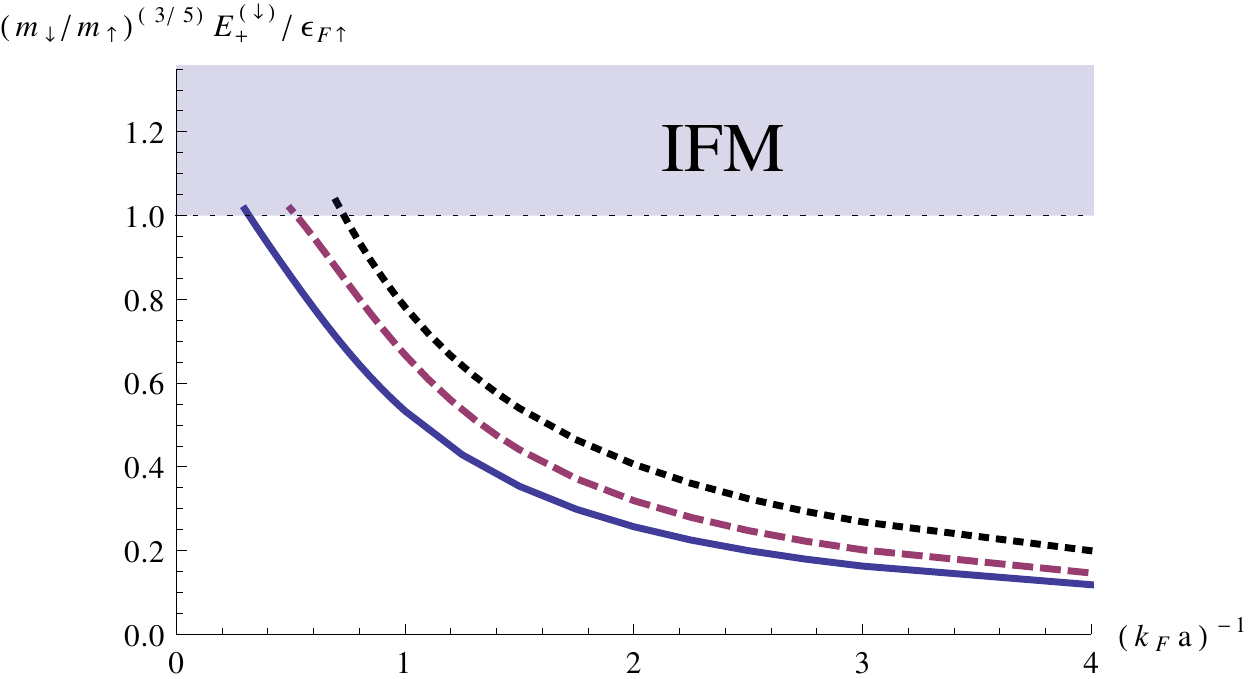}
\end{center}
\caption{The stability of the IFM phase is determined by the simultaneous satisfaction of the two conditions in Eq.~(\ref{IFMconditions}). The relevant adimensional quantity is plotted here. From top to bottom, lines correspond to $m_\downarrow/m_\uparrow=40/6,\,6/40,\,1$.}
\label{figIFM}
\end{figure}
%%%%%%%%%%%%%%%%%%%%%%%%%%%%%%%%%%%%%%%%%%%%%%%%%%%%%%%%%%%%%%%%%%%%%%%%
The completely phase separated state is favored only if both these two conditions are satisfied. They are 
plotted in Fig.~\ref{figIFM} for various mass ratios. 
For equal masses, we find that a fully ferromagnetic phase is favored for $0<1/k_Fa\lesssim 0.4$,
 in agreement with the result in Ref.~\cite{Cui10}.

The mass imbalanced case is slightly more complicated. As an example, let us consider the  conditions for a mixture of $^6$Li and $^{40}$K atoms to fully  separate
into a region of K atoms with Fermi energy $\epsilon_{F,{\rm K}}$ and a region of Li atoms with Fermi energy $\epsilon_{F,{\rm Li}}$. 
The condition for the stability against K atoms to move across the phase boundary is $E_+^{\rm K}/\epsilon_{F,{\rm Li}}> (6/40)^{3/5}$.
From Fig.~\ref{figIFM}, we see that this condition is fulfilled for  $1/k_{F,{\rm Li}}a\lesssim0.7$.
Likewise, the stability against Li atoms moving across the phase boundary is $E_+^{\rm Li}/\epsilon_{F,{\rm K}}> (40/6)^{3/5}$ which 
gives $1/k_{F,{\rm K}}a\lesssim0.5$. Since $k_{F,{\rm K}}/k_{F,{\rm Li}}=(40/6)^{1/5}\simeq1.5$, these two conditions essentially give the same critical value for the 
scattering length. Interestingly, the stable region for the ferromagnetic phase seems to be larger than for the equal mass case.

It is however important to consider the lifetime of the repulsive polarons in this argument. As discussed above, 
 qrepulsive polaron become increasingly unstable towards decay as the coupling strength increases.
The characteristic energy scale associated to the many-body effects is the majority Fermi energy, which is of order 10-50kHz in current experiments. Thus, in order to observe the formation of ferromagnetic domains one should require $\Gamma/\epsilon_F \ll 1$ in addition to Eq.\ (\ref{IFMconditions}).
 For the case of equal masses, we find $\Gamma/\epsilon_F \sim 0.5$ at the critical interaction $1/k_Fa=0.4$, showing that the gas is strongly unstable towards decay into lower-lying excitations, in accord with the results of \cite{Pekker10} for a balanced gas. Also for the Li-K mixture considered above, both decay rates of the heavy (K) and light impurities (Li) are of order the Fermi energy at the critical strength, signaling a strong instability of the IFM phase towards decay into lower excitations  for unequal masses.

%%%%  %%%%  %%%%  %%%%

\section{Spectral functions and rf spectrum}
The single particle properties of the impurity can be probed with radio-frequency (rf)  spectroscopy~\cite{Chin04,Shin07,Stewart08}. 
One can either flip the spin state of the impurity  to a third state or flip a  third spin state to the impurity state by a rf photon.
A clearcut theoretical interpretation of rf experiments is in general hard since effects such as self-energies, pairing, and
trap inhomogeneities have to be included. If there are significant interactions with the third state, the interpretation is complicated  further by vertex corrections~\cite{Perali,Pieri,He}. 
In the limit of a  large impurity mass, analytical calculations indeed show that vertex corrections  can
 change the rf spectrum qualitatively~\cite{BruunClock}. 
 
The theoretical analysis of rf spectroscopy is simplified considerably when there are only significant effects of the interactions between the impurity 
and the majority Fermi sea. In this case, the rf spectrum is determined by a sum over the spectral functions of the occupied states for the 
impurity~\cite{TwinPeaks,Massignan08}. In Fig.\ \ref{figRFspectra}, we plot the spectral function (\ref{Spectral})
of the impurity for $m_\downarrow/m_\uparrow=1$ as a function of interaction 
strength for two different momenta. These plots illustrate that the spectral weight of the quasiparticle peak of the 
attractive polaron decreases as the interaction is changed from the 
BCS to the BEC side of the resonance. The repulsive polaron on the other hand is only well defined on the BEC side. It could be interesting to see 
this transition of spectral weight from the attractive to the repulsive polaron quasiparticle peak with increasing interaction. 
It is presumably most easily done by using the non-interacting third state as an initial state so that the rf photon flips it into the impurity state. 
In this way one avoids complications due to thermal depletion and decay of the repulsive polaron branch.

%%%%%%%%%%%%%%%%%%%%%%%%%%%%%%%%%%%%%%%%%%%%%%%%%%%%%%%%%%%%%%%%%%%%%%%%
\begin{figure}
\begin{center}
\includegraphics[width=\columnwidth]{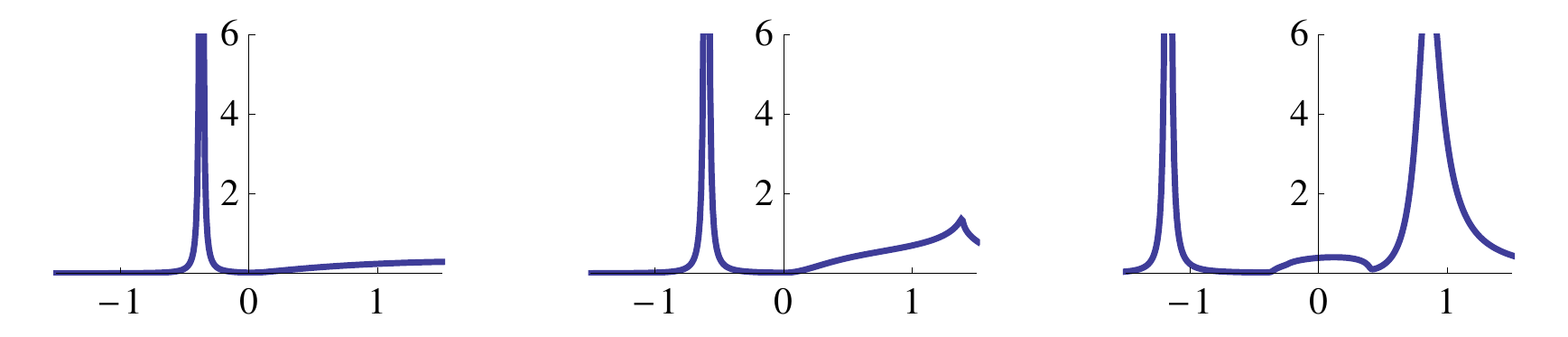}
\includegraphics[width=\columnwidth]{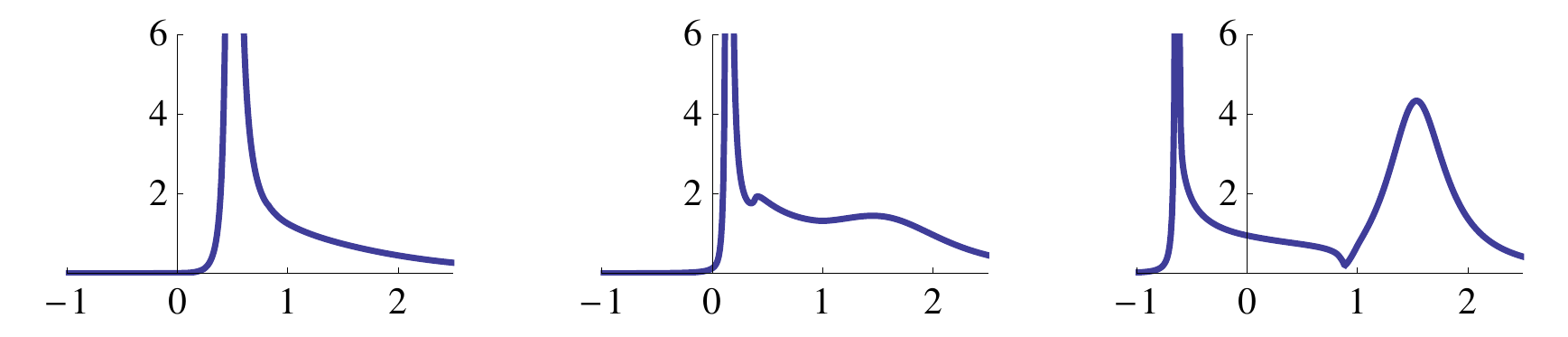}
\end{center}
\caption{The spectral function $A_\downarrow(\bk,\omega)$ for the mass balanced case. Top row: $|\bp|/k_F=0.1$.
 Bottom row $|\bp|/k_F=1$.
 From left to right: $1/k_Fa=-0.5,0,0.5$.
The x-axis is the energy $\omega/\epsilon_F$.
}
\label{figRFspectra}
\end{figure}
%%%%%%%%%%%%%%%%%%%%%%%%%%%%%%%%%%%%%%%%%%%%%%%%%%%%%%%%%%%%%%%%%%%%%%%%

%%%%  %%%%  %%%%  %%%%

\section{Conclusions}
We analyzed 
the metastable repulsive polaronic branch which appears at positive energies and scattering lengths for the case of an impurity in an ideal Fermi gas. The polaron 
 energy, decay rate, effective mass, quasiparticle residue, and spectral function were calculated as a function of interaction strength and mass ratio.
  Those properties may be  probed by radio-frequency spectroscopy.
   We used these results to derive conditions for perfect ferromagnetism for a binary mixture of atoms with a general mass ratio. Our results 
   show that the transition towards itinerant ferromagnetism should happen at a weaker interaction strength in a mixture of unequal masses. However,
     short quasiparticle lifetimes seem to prevent the experimental observation of itinerant ferromagnetism in universal Fermi-Fermi mixtures.
 % We finally plotted the spectral function for the polaron for various momenta and interaction strengths, and we discussed how it can be probed using radio-frequency spectroscopy.

\begin{acknowledgement}
We wish to thank Matteo Zaccanti, Hui Zhai, Sascha Z\"ollner, Sebastiano Pilati, and Maciej Lewenstein for stimulating discussions.
P.M.\ acknowledges funding from ESF project FERMIX (FIS2007-29996-E), ERC Advanced Grant QUAGATUA, and Spanish MEC projects TOQATA (FIS2008-00784) and FIS2008-01236. This work is a result of a collaboration which originated
within the POLATOM Research Networking Programme of the
European Science Foundation (ESF).
Part of this work was done while one of the authors (P.M.) participated in the program BOPTILATT at the Kavli Institute for Theoretical Physics. This research was supported in part by the National Science Foundation under Grant No. NSF PHY05-51164.
\end{acknowledgement}


\begin{thebibliography}{10}
\bibitem{RevMods} S.\ Giorgini, L.\ P.\ Pitaevskii, S.\ Stringari, Rev.\ Mod.\ Phys. \textbf{80}, (2008) 1215;
I.\ Bloch, J.\ Dalibard, W.\ Zwerger, Rev.\ Mod.\ Phys. \textbf{80}, (2008) 885.
\bibitem{PolaronPapers}F.\ Chevy and C.\ Mora, Rep.\ Prog.\ Phys. \textbf{73}, (2010) 112401. 
\bibitem{Taglieber08} M. Taglieber {\it et al.}, Phys. Rev. Lett. {\bf 100}, (2008) 010401.
\bibitem{Wille08}E. Wille {\it et al.},
% F. M. Spiegelhalder, G. Kerner, D. Naik, A. Trenkwalder, G. Hendl, F. Schreck, R. Grimm, T. G. Tiecke, J. T. M. Walraven, S. J. J. M. F. Kokkelmans, E. Tiesinga, and P. S. Julienne, Exploring an Ultracold Fermi-Fermi Mixture: Interspecies Feshbach Resonances and Scattering Properties of 6 Li and 40 K,
Phys. Rev. Lett. {\bf 100}, (2008) 053201.
\bibitem{Tiecke10} T. G. Tiecke {\it et al.}, Phys. Rev. Lett. {\bf 104}, (2010) 053202.
\bibitem{Naik10} D. Naik {\it et al.},
% Feshbach resonances in the 6Li-40K Fermi-Fermi mixture: Elastic versus inelastic interactions,
 arXiv:1010.3662.
 \bibitem{Fukuhara07}T. Fukuhara, Y. Takasu, M. Kumakura, and Y. Takahashi,
%Degenerate Fermi Gases of Ytterbium,
   %173Yb
 Phys. Rev. Lett. {\bf 98}, (2007) 030401.
\bibitem{Schirotzek09} A. Schirotzek, C.-H. Wu, A. Sommer, and M. W. Zwierlein, Phys. Rev. Lett. {\bf 102}, (2009) 230402.
\bibitem{Zoellner} S.\ Z\"ollner, G.\ M.\ Bruun, and  C.\ J.\ Pethick, Phys. Rev. A {\bf 83}, (2011) 021603(R).
\bibitem{Parish11} M. Parish, Phys. Rev. A {\bf 83}, (2011) 051603(R).
\bibitem{Giraud09} S.\ Giraud and R.\ Combescot, Phys. Rev. A {\bf 79}, (2009) 043615.
\bibitem{Leskinen10} M. J. Leskinen, O. H. T. Nummi, F. Massel and P. T\"orm\"a, New J. Phys. {\bf 12}, (2010) 073044.
\bibitem{Stoner33} E. Stoner, Phil. Mag. {\bf 15}, (1933) 1018.
\bibitem{Jo} G.-B. Jo  {\it et al.}, Science \textbf{325}, (2009) 1521.
%\bibitem{Conduit}G.\ J.\ Conduit, A.\ G.\ Green, and B.\ D.\ Simons, Phys. Rev. Lett. {\bf 103}, (2009) 207201.
\bibitem{Zhai09} H. Zhai,
%Correlated versus Ferromagnetic State in Repulsively Interacting Two-component Fermi Gases
Phys. Rev. A \textbf{80}, (2009) 051605(R).
\bibitem{Cui10} X. Cui and H. Zhai, Phys. Rev. A {\bf 81}, (2010) 041602(R).
\bibitem{Pilati10} S. Pilati, G. Bertaina, S. Giorgini, and M. Troyer, Phys. Rev. Lett. {\bf 105}, (2010) 030405.
\bibitem{Pekker10} D. Pekker {\it et al.}, Phys. Rev. Lett. \textbf{106}, (2011) 050402.
\bibitem{Chevy}F.\ Chevy, Phys. Rev. A {\bf 74}, (2006) 063628.
\bibitem{Prokofev08} N. Prokof'ev and B. Svistunov, Phys. Rev. B \textbf{77}, (2008) 020408(R); \textbf{77}, (2008) 125101.
\bibitem{Combescot07} R. Combescot, A. Recati, C. Lobo, and F. Chevy, Phys. Rev. Lett. \textbf{98}, (2007) 180402.
%\bibitem{MahanBook} G. D. Mahan, {\it Many-particle physics} (Springer Verlag 2008).
\bibitem{Punk} C. Mora and F. Chevy, Phys. Rev. A {\bf 80}, (2009) 033607;
M.\ Punk, P.\ T.\ Dumitrescu, and W.\ Zwerger, Phys. Rev. A {\bf 80}, (2009) 053605;
R.\ Combescot, S.\ Giraud, and X.\ Leyronas, Europhys.\ Lett.\ \textbf{88}, (2009) 600007.
\bibitem{Petrov03}D. S. Petrov, Phys. Rev. A {\bf 67}, (2003) 010703(R).
\bibitem{Bruun10} G. M. Bruun and P. Massignan, Phys. Rev. Lett.\ \textbf{105}, (2010) 020403.
\bibitem{Fumi} G. D. Mahan, Many-Particle Physics, 2nd ed. (Plenum, New York, 1990), p. 253; A. Collin, P. Massignan, C. J. Pethick, Phys. Rev. A {\bf 75}, (2007) 013615.
\bibitem{Bishop73} R. F. Bishop, Ann. Phys. \textbf{78}, (1973) 391.
%\bibitem{Okano10} M. Okano {\it et al.},Appl. Phys. B {\bf 98}, (2010) 691.

% H. Hara, M. Muramatsu, K. Doi, S. Uetake, Y. Takasu and Y. Takahashi,
%Simultaneous magneto-optical trapping of lithium and ytterbium atoms towards production of ultracold polar molecules,
   %6Li-174Yb   (bosonic Ytterbium and fermionic Lithium)
\bibitem{Massignan05} P. Massignan, C. J. Pethick, and H. Smith, Phys. Rev. A {\bf 71}, (2005) 023606.
\bibitem{Sadeghzadeh11} K. Sadeghzadeh, G. Bruun, C. Lobo, P. Massignan, and A. Recati, arXiv:1012.0484 (NJP in press).
\bibitem{Duine05}R. A. Duine and A. H. MacDonald, Phys. Rev. Lett. {\bf 95}, (2005) 230403.
\bibitem{Keyserlingk}C.\ W.\ von Keyserlingk and G.\ J.\ Conduit, arXiv:11041439.
\bibitem{Chin04} C. Chin {\it et al.},
%M. Bartenstein, A. Altmeyer, S. Riedl, S. Jochim, J. Hecker Denschlag, R. Grimm
%"Observation of the Pairing Gap in a Strongly Interacting Fermi Gas", 
Science {\bf 305}, (2004) 1128.
\bibitem{Shin07} Y. Shin, C. H. Schunck, A. Schirotzek,  W. Ketterle,
%"Tomographic rf Spectroscopy of a Trapped Fermi Gas at Unitarity",
 Phys. Rev. Lett. {\bf 99}, (2007) 090403.
\bibitem{Stewart08} J. T. Stewart, J. P. Gaebler,  D. S. Jin,
%"Using photoemission spectroscopy to probe a strongly interacting Fermi gas",
Nature {\bf 454}, (2008) 744.
\bibitem{Perali}A.\ Perali, P.\ Pieri, and G.\ C.\ Strinati, Phys.\ Rev.\ Lett.\ {\bf 100}, (2008) 010402.
\bibitem{Pieri}P.\ Pieri, A.\ Perali, and  G.\ C.\ Strinati, Nat.\ Phys.\ \textbf{5}, (2009) 736.
\bibitem{He}Y.\ He, C.-C.\ Chien, Q.\ Chen, and K.\ Levin, Phys.\ Rev.\ Lett.\ {\bf 102}, (2009) 020402.
\bibitem{BruunClock}G.\ M.\ Bruun, C.\ J.\ Pethick, and Z.\ Yu, Phys. Rev. A {\bf 81}, (2010) 033621.
\bibitem{TwinPeaks} P. Massignan, G. M. Bruun, and H. T. C. Stoof, Phys. Rev. A {\bf 77}, (2008) 031601.
\bibitem{Massignan08} P. Massignan, G. M. Bruun, and H. T. C. Stoof, Phys. Rev. A {\bf 78}, (2008) 031602(R).

\end{thebibliography}
\end{document}